\documentclass[conference]{IEEEtran}
\IEEEoverridecommandlockouts
\usepackage{cite}
\usepackage{amsmath,amssymb,amsfonts}
\usepackage{algorithmic}
\usepackage{graphicx}
\usepackage{textcomp}
\usepackage{xcolor}
\def\BibTeX{{\rm B\kern-.05em{\sc i\kern-.025em b}\kern-.08em
T\kern-.1667em\lower.7ex\hbox{E}\kern-.125emX}}

\usepackage{amsmath,amsfonts}

\usepackage{algorithmic}

\usepackage{algorithm}

\usepackage{array}

\usepackage[caption=false,font=normalsize,labelfont=sf,textfont=sf]{subfig}

\usepackage{textcomp}

\usepackage{stfloats}

\usepackage{url}

\usepackage{verbatim}

\usepackage{graphicx}

\usepackage{cite}
 
\usepackage{xcolor}
\usepackage{multirow}

\usepackage{soul}    
\begin{document}

\title{Reference-Free EM Validation Flow for Detecting Triggered Hardware Trojans}

\author{\IEEEauthorblockN{Mahsa Tahghigh}
\IEEEauthorblockA{{mahsa.tahghigh@bison.howard.edu} \\
\textit{Howard University}\\ Washington, D.C, USA}
\and
\IEEEauthorblockN{Hassan Salmani}
\IEEEauthorblockA{{hassan.salmani.howard.edu} \\
\textit{Howard University}\\ Washington, D.C, USA }}

\maketitle

\begin{abstract}
Hardware Trojans (HTs) threaten the trust and reliability of integrated circuits (ICs), particularly when triggered HTs remain dormant during standard testing and activate only under rare conditions. Existing electromagnetic (EM) side-channel–based detection techniques often rely on golden references or labeled data, which are infeasible in modern distributed manufacturing. This paper introduces a reference-free, design-agnostic framework for detecting triggered HTs directly from post-silicon EM emissions. The proposed flow converts each EM trace into a time–frequency scalogram using Continuous Wavelet Transform (CWT), extracts discriminative features through a convolutional neural network (CNN), reduces dimensionality with principal component analysis (PCA), and applies Bayesian Gaussian Mixture Modeling (BGMM) for unsupervised probabilistic clustering. The framework quantifies detection confidence using posterior-based metrics ($\alpha_{post}$, $\beta_{post}$), Bayesian information criterion ($\Delta BIC$), and Mahalanobis cluster separation ($D$), enabling interpretable anomaly decisions without golden data. Experimental validation on AES-128 designs embedded with four different HTs demonstrates high separability between HT-free and HT-activated conditions and robustness to PCA variance thresholds. The results highlight the method’s scalability, statistical interpretability, and potential for extension to runtime and in-field HT monitoring in trusted microelectronics.
\end{abstract}


\section{Introduction}
Microelectronics power nearly every sector of modern society, from consumer electronics and cloud services to defense and mission-critical systems \cite{chesebrough2017trusted}\cite{bis2023microelectronics}. At the heart of these devices are complex System-on-Chip (SoC) architectures, whose design and fabrication are increasingly outsourced to offshore foundries\cite{pan2022survey}\cite{jain2021survey}. While outsourcing reduces cost and accelerates production, it introduces serious supply-chain security risks\cite{hanindhito2019hardware}\cite{vashistha2018trojan}\cite{9310331}.

A major concern is the insertion of hardware Trojans (HTs), which can alter functionality, leak sensitive data, or enable backdoor access \cite{Xiao2016Hardware}\cite{tahghigh2024gmm}\cite{11014346}\cite{TahghighArXiv2026}. The most challenging class are triggered HTs, which remain dormant during normal operation and activate only under rare conditions, evading traditional testing and verification\cite{ghimire2023quantum}. 
Prior circuit-level studies have shown that post-fabrication hardware behavior may diverge from design-time assumptions, complicating validation in deployed systems and motivating reference-free analysis
methods \cite{cta3980}. Detection is further hindered by the absence of golden references (trusted designs or known-good chips), making many existing detection techniques impractical\cite{he2017chipfree}.

Prior HT detection methods often rely on golden-reference comparisons, using functional or side-channel tests to flag deviations. These approaches are effective in controlled environments but unrealistic in modern supply chains. Other efforts leverage electromagnetic (EM) side-channel analysis with supervised machine learning, which requires labeled datasets or knowledge of HT behavior\cite{jap2016supervised}\cite{8952724}. While promising, such methods cannot scale to unknown or stealthy triggered HTs\cite{gubbi2023tutorial}\cite{gubbi2022survey}.

This paper introduces a reference-free EM detection flow designed for post-silicon validation without requiring golden models, design details, or manufacturing information. The flow applies Continuous Wavelet Transform (CWT) to individual EM traces, generating rich time–frequency representations. A convolutional neural network (CNN) extracts discriminative features, which are reduced using principal component analysis (PCA) and clustered with a Bayesian Gaussian Mixture Model (BGMM) for unsupervised anomaly detection. Unlike prior EM-based approaches, this method is design-agnostic, unsupervised, and tailored to capture the rare activation patterns of triggered HTs.

This work:
\begin{itemize}
    \item Proposes a reference-free EM detection methodology requiring no golden models or labeled training.
    \item Leverages CWT-based per-trace analysis to capture transient anomalies.
    \item Integrates CNN, PCA, and BGMM into an unsupervised probabilistic framework that yields detection confidence metrics.
    \item Demonstrates a design-agnostic, post-silicon flow suitable for scalable deployment in supply-chain assurance.
\end{itemize}

This post-silicon, design-agnostic approach provides a scalable and practical framework for hardware trust. It directly targets triggered HTs—the most challenging class of threats—and advances the state of the art by combining time–frequency analysis, deep feature learning, and probabilistic anomaly modeling into a unified detection flow suitable for supply-chain assurance and validation workflows.

The remainder of this paper is organized as follows. Section~\ref{TheProposedTechnique} details the proposed detection technique. Section~\ref{ExperimentalSetup} describes the experimental setup and measurement process. Section~\ref{ExperimentalEvaluationsandResults} presents the experimental results and analyses. Section~\ref{RelatedWork} reviews relevant prior work. Finally, Section~\ref{Conclusion} concludes the paper and outlines future research directions.

\begin{figure*}[t]
    \centering
    \includegraphics[width=0.70\linewidth, angle=360]{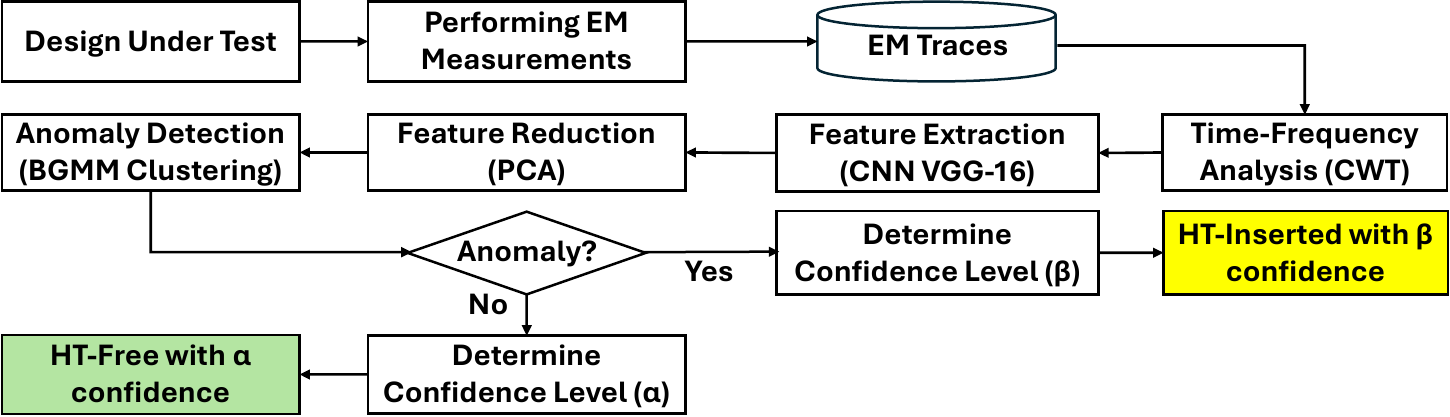}
    \caption{EM-Based Detection flow for Triggered Hardware Trojans.}
    \vspace{-0.15in}
    \label{fig:EMHTDetFlow}
\end{figure*} 

\section{The Proposed Technique}
\label{TheProposedTechnique}
Figure \ref{fig:EMHTDetFlow} presents the proposed HT detection flow 
for detecting triggered HTs. 
We operate in a post-silicon, black-box setting: 
the device is already manufactured; 
no netlist, layout, or fab metadata is available; 
and no golden reference or labeled data is assumed. 
The core challenge is to surface rare, short-lived anomalies 
in EM emissions when a triggered HT activates.
Our flow couples time–frequency analysis with 
unsupervised probabilistic modeling to convert raw EM traces 
into confidence-rated decisions without design knowledge.

\begin{figure}[t]
    \centering
    \includegraphics[width=.7\linewidth]{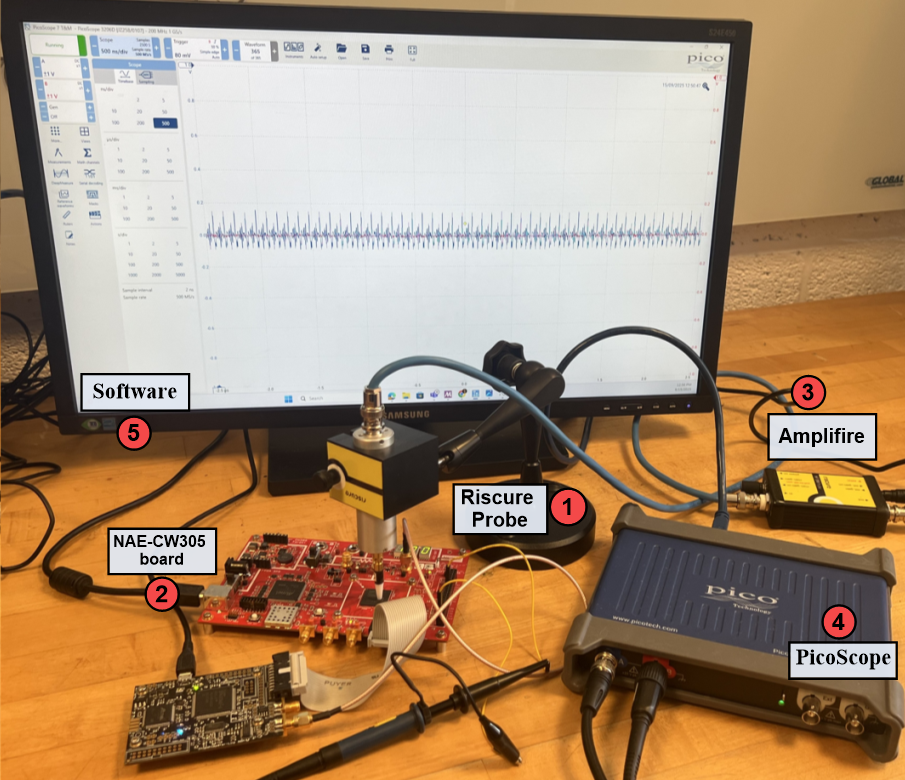}
    \vspace{-.1in}
    \caption{Experimental Setup.}
    \vspace{-.3in}
    \label{fig:ExperimentalSetup}
    
\end{figure}   

\textit{A. Threat Model and Operating Assumptions:} Triggered HTs represent one of the most challenging microelectronic threats, remaining dormant under normal conditions and activating under rare input or state sequences—producing transient EM perturbations. Our framework targets this scenario in a post-silicon, black-box environment that mirrors modern distributed manufacturing and trust-assurance workflows.

1. No Design Knowledge or Golden Reference: No access to layout, netlist, or labeled reference data is assumed. The method must infer trustworthiness directly from the observed EM behavior, without any golden model.

2. Statistical Trace Acquisition: EM traces are collected while stimulating the DUT through structured but black-box workloads to increase the probability of HT activation. Since triggered HTs may appear in only 2–5\% of traces, the analysis emphasizes statistical sufficiency over deterministic test coverage.

3. Noise and Environmental Variability: The technique remains robust under session-to-session and environmental variations, tolerating realistic measurement noise.

Operating under these constraints, the proposed reference-free and design-agnostic framework addresses key supply-chain trust challenges. Its sequential EM processing also enables scalable runtime or in-field HT detection for continuous assurance.

\textit{B. EM Trace Acquisition and Pre-Processing}: We collect EM traces while running structured but black-box stimuli (functional tests or workload scripts). Each trace is normalized (e.g., zero-mean/unit-variance or per-trace energy normalization) and optionally band-limited to remove out-of-band noise. When possible, we stabilize sampling rate and probe placement; session offsets are mitigated by per-trace normalization. Triggered HTs manifest as rare events. High-throughput acquisition and trace normalization improve the signal-to-anomaly ratio and reduce false flags due to session drift or probe repositioning.

\textit{C. Per-Trace Time–Frequency Representation via CWT}: Each trace is transformed by CWT to produce a scalogram (time–frequency energy map). We use a complex Morlet wavelet (or similar) for good time-frequency localization; scales are chosen to cover bands where digital switching and clock harmonics reside. CWT adapts its time–frequency resolution with scale, making it well suited to short, sparse, and nonstationary HT activations. A scalogram is a 2D representation where when and at what frequencies energy deviates is explicitly visible—ideal for downstream learning. Applying CWT trace-by-trace avoids window-sweep heuristics and keeps the method lightweight and consistent across designs.

\begin{figure}[!t]
\centering
\subfloat[]{\includegraphics[width=1.5in]{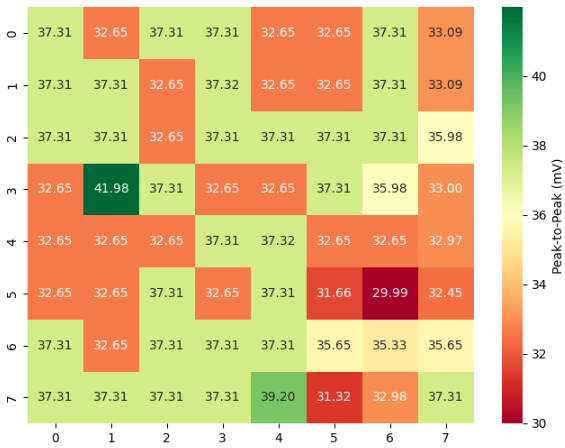}}
\hfil
\subfloat[]{\includegraphics[width=1.5in]{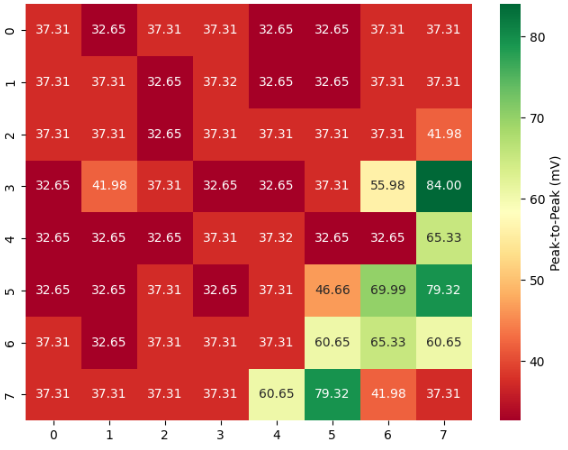}}
\caption{$8 \times 8$ heatmaps displaying the peak-to-peak difference in EM signals. (a) illustrates the baseline signal without AES, while (b) reveals the impact of AES implementation on signal amplitude.}
\vspace{-.2in}
\label{fig:heatmap}
\end{figure} 

\textit{D. Deep Feature Extraction (CNN) and Dimensionality Reduction (PCA)}: 
The objective of feature extraction is to transform the time–frequency scalograms into a low-dimensional vector space where the subtle, nonstationary characteristics induced by triggered HTs are maximally separable from the normal background activity. Hand-crafted features often struggle with subtle, layout-specific EM patterns and lack the generalization required for a design-agnostic solution.

We address this by feeding the CWT scalograms to a CNN, specifically utilizing a VGG-16 backbone. CNNs are highly effective at learning discriminative structures—such as localized bursts, harmonics, and side-band textures—that characterize the transient EM perturbation caused by HT activation.
To maintain the reference-free and unsupervised nature of our flow while ensuring feature quality, we leverage transfer learning:

\begin{figure*}[t]
    \centering
    \includegraphics[width=.80\linewidth]{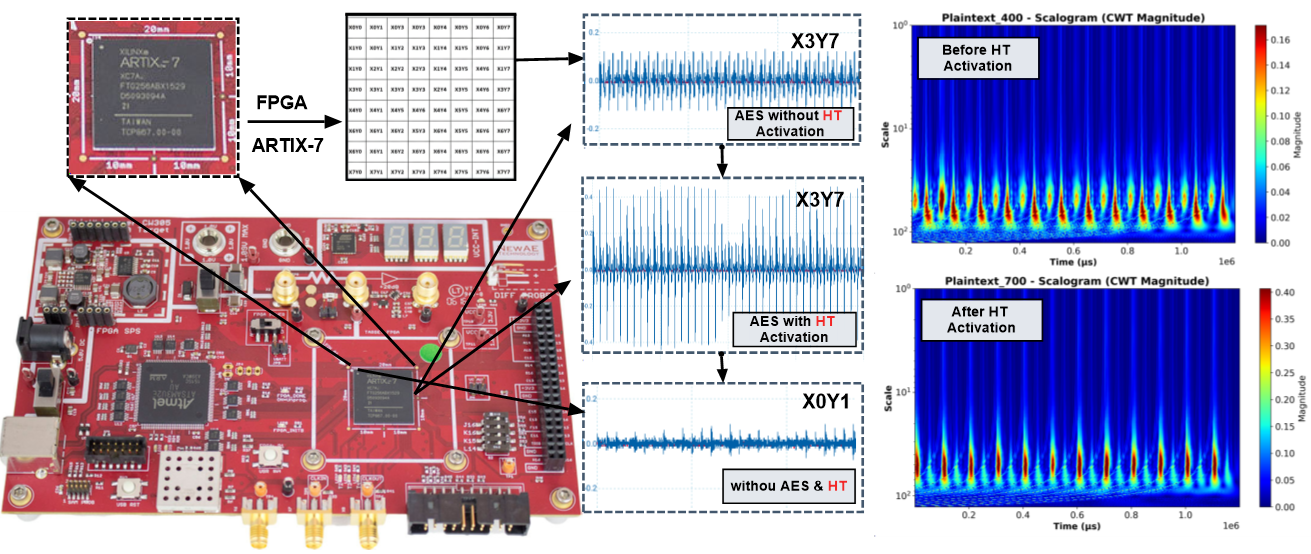}
    \vspace{-0.1in}
    \caption{Representative EM traces and corresponding time–frequency scalograms captured from FPGA regions with and without AES activity. The weak signal from inactive grid X0Y1 contrasts with the stronger emissions from the AES region X3Y7. The scalograms illustrate how hardware Trojan (HT\#1) activation alters the time–frequency structure of the EM signal, changing both its energy distribution and waveform morphology.}
    \vspace{-.2in}
    \label{fig:SampleEMs}
\end{figure*}

1. Feature Initialization: Early layers of the VGG-16 backbone are typically frozen because they learn generic edge and texture filters. These serve as a robust, pre-initialized basis for feature extraction, allowing the network to recognize basic structural patterns inherent in any 2D data representation, including time–frequency scalograms.
2. Specialization: We specialize the higher-level feature extraction by optionally fine-tuning late layers on unlabeled scalograms using self-supervised pretext tasks (e.g., rotation prediction) or light contrastive augmentation. This ensures the extracted features are tailored to the sparse, transient EM patterns of the trigger events, keeping the feature pipeline fully unsupervised.

The penultimate layer of the specialized CNN yields 
a high-dimensional $512$ feature vector of $7 \times 7$ 
convolutional feature map capturing 
these salient time–frequency patterns. 
To reduce feature dimensionality and avoid overfitting risk, 
a global average 2D pooling operation is applied to 
the final feature maps.
This operation effectively condenses 
25,088 spatial activations $512 \times 7 \times 7$ into 512 global descriptors, 
capturing the most dominant and spatially invariant characteristics of the input.
Following feature extraction, we apply PCA to the resultant feature vectors:
1. Dimensionality Management: PCA is used to avoid the curse of dimensionality in the subsequent clustering step.
2. Noise Mitigation: It helps suppress redundant and noisy directions in the feature space.
3. Stability: PCA is critical for improving the stability of the mixture model clustering.

In practice, we retain components explaining 90\% of 
the total variance (this threshold is tunable). 
To obtain a compact representation for each sample (i.e., an EM trace), 
we compute the normalized weighted sum of its retained principal component scores, 
where each component is scaled by its corresponding explained variance ratio, that is \[Sample_i = w_1 \cdot PC_{i,1} + w_2 \cdot PC_{i,2} + \cdots + w_n \cdot PC_{i,n}\]
where \( PC_{i,k} \) denotes the value of the \( k^{\text{th}} \) principal component 
for sample \( i \), and \( w_k \) represents its variance contribution. 
This formulation emphasizes the most informative directions 
while preserving the overall variance structure, 
resulting in a stable one-dimensional descriptor for each sample.

The combined CNN-PCA pipeline provides an unsupervised, stable, and statistically sound set of features, tailored to trigger transients, which is essential for the subsequent probabilistic modeling.

\begin{table*}[t]
    \centering
    \begin{tabular}{|c|c|c|c|}\hline
         \textbf{Name}&  \textbf{Attack Type}& \textbf{Trigger}& \textbf{Payload}\\\hline
         HT\#1 & Leakage Information & Specific sequence of & Encryption key bits\\\hline
         HT\#2 & \multicolumn{3}{|c|}{Augmented version of HT\#1}\\\hline
         HT\#3 & Denial of Service & A specific input plaintext & Encryption disabled\\\hline
         HT\#4 & Denial of Service & Any plaintext input from a specific set of values & Encryption disabled\\\hline
    \end{tabular}
    \caption{The characteristics of HTs inserted in AES-128}
    \vspace{-.4in}
    \label{tlb:HTsCharacter}
\end{table*}

\textit{E. Unsupervised Probabilistic Modeling with BGMM}: Reduced features are modeled by a BGMM. Using Bayesian, Dirichlet priors prune unnecessary components, preventing overfitting when anomalies are sparse, and giving calibrated mixture weights. Furthermore, The BGMM yields (i) mixture weights (how much mass each cluster has), (ii) posterior responsibilities per trace, and (iii) sample likelihoods.

\textit{F. Anomaly Determination Prior to Confidence Calculation}: While the BGMM produces clusters 
from EM-trace features, not every multi-cluster result should be interpreted as an anomaly. 
Gaussian mixtures may split a single population into 
several overlapping components for better fit, even in HT-free devices. 
To avoid false alarms, we first perform anomaly validation before computing 
the confidence metrics ($\alpha$, $\beta$):

\begin{itemize}
    \item \textbf{Identify the dominant cluster} The cluster with the largest mixture weight ($\pi_{max}$) reported by BGMM is labeled as the "normal" cluster, reflecting the majority of traces.
    \item \textbf{Evaluate secondary clusters} Each additional cluster is examined using two criteria:
    \begin{itemize}
        \item \textbf{Mixture weight threshold:} Clusters representing less than 2–5\% of total traces are treated as spurious noise fits.
        \item \textbf{Cluster separation:} Bhattacharyya or Mahalanobis distance is used to measure how distinct a secondary cluster is from the dominant one. Overlapping or weakly separated clusters are ignored.
    \end{itemize}
    \item \textbf{Decision:}
    \begin{itemize}
        \item    If no secondary cluster passes both thresholds, the dataset is declared HT-free, and we report $\alpha \approx 1$, $\beta \approx 0$.
        \item    If at least one credible anomaly cluster exists, the dataset is flagged as HT-suspect. In this case, we proceed with computing $\alpha$ and $\beta$ from posterior probabilities, providing quantitative confidence levels of normal vs. HT behavior.
    \end{itemize}        
\end{itemize}

If no secondary cluster passes the anomaly validation checks, the dataset is classified as HT-free. Otherwise, we proceed to quantify the detection confidence using $\alpha$ (normal behavior) 
and $\beta$ (anomalous behavior). In the following, we define their posterior-based realizations, $\alpha_{post}$ and $\beta_{post}$, derived from the BGMM.

\textit{G. Confidence Level Calculation}: Once a credible anomaly cluster is identified, we quantify the likelihood that the circuit is HT-free or HT-suspect using posterior-based confidence metrics. For each trace, the BGMM assigns posterior probabilities across clusters. Let $k^*$ denote the dominant (normal) cluster.

\begin{itemize}
    \item \textbf{Posterior confidence per trace:}\\
    $p_i^{norm}=r_{i,k^*},\;\;\;    p_i^{anorm}= 1 - p_i^{norm}$\\
    where $r_{i,k^*}$ is the posterior probability that trace \textit{i} belongs to the dominant cluster.\\
    It is noted that\\ 
    $r_{i,k} = \frac{\pi_i f_k (x_i)}{\sum_{j}\pi_j f_j (x_i)}$\\
    where $f_k (x)$ is the density of component $k$ at $x_i$, and $\pi_k$ is its weight. 
    \item \textbf{Aggregate confidence across all traces:}\\
    $\alpha_{post} = \frac{1}{N}\sum_{i=1}^{N}p_i^{norm}$\;\; $\beta_{post} = \frac{1}{N}\sum_{i=1}^{N}p_i^{anorm}$\\
    here $\alpha_{post}$ reflects the confidence level that the majority of traces are consistent with normal behavior, while $\beta_{post}$ measures the prevalence of anomalous behavior. 

    \item \textbf{Decision rule}: We declare a device HT-suspect using a two-evidence criterion: (i) model evidence of multi-modality via $\Delta \mathrm{BIC}$ and (ii) anomaly prevalence via $\beta_{\text{post}}$. Geometric separation $D$ is reported as supporting context.\\    
    \noindent\textbf{High confidence:}
    \[
    \Delta \mathrm{BIC} \ge 10 ~~ and ~~ \beta_{\text{post}} \ge 0.30.
    \]
    \noindent\textbf{Moderate confidence:}
    \[
    \Delta \mathrm{BIC} \ge 10 ~~ and ~~ 0.20 \le \beta_{\text{post}} < 0.30,
    \]
    with either $D \ge \sqrt{\chi^2_{p,\,0.99}}$ or consistent outcomes across PCA thresholds/windows.\\
    \noindent\textbf{Not suspicious:}
    \[
    \Delta \mathrm{BIC} < 10 ~~ or ~~ \beta_{\text{post}} < 0.20.
    \]
    \noindent
    For interpretability, we report $D$; in two-dimensional views $D \gtrsim 3$ approximates the $99\%$ $\chi^2$ contour \cite{Etherington2019}. We treat $D$ as a \emph{soft} indicator—sensitive to transient or low-duty-cycle activations—while $\Delta \mathrm{BIC}$ and $\beta_{\text{post}}$ provide primary statistical evidence. 
\end{itemize}

This probabilistic approach avoids binary judgments and instead provides a quantitative confidence measure of tampering. 

\section{Experimental Setup}
\label{ExperimentalSetup}
The experimental setup, shown in Fig. \ref{fig:ExperimentalSetup}, emulates a post-silicon trust verification environment for detecting HTs. It comprises a ChipWhisperer NAE-CW305 FPGA platform running the target design, coupled to a Riscure EM measurement station equipped with a 1.25 mm EM probe to capture side-channel emissions during circuit operation. Signals are digitized using a PicoScope 3000-series oscilloscope at a 500 MS/s sampling rate, with captured waveforms exported as CSV files for subsequent time–frequency analysis and anomaly detection.

In this paper, we evaluate the proposed detection flow for triggered HTs on an AES-128 encryption core modified with four representative HTs, summarized in Table \ref{tlb:HTsCharacter}. Two distinct HT families are used as primary detection targets. HT\#1 and its variant HT\#2 are derived from the AES-T1100 HT in Trust-Hub. These HTs activate only when a specific plaintext pattern is applied, after which they covertly leak the secret AES key through a hidden communication channel. Inspired by spread-spectrum techniques such as Code-Division Multiple Access (CDMA), the leakage is dispersed across several clock cycles: a pseudo-random number generator (PRNG), seeded with the triggering plaintext, produces a CDMA code that XOR-modulates the secret bits. The modulated sequence then drives a leakage circuit (LC) built from eight flip-flops attached to the XOR output, emulating large capacitance and creating a covert, CDMA-based side channel. 

Based on the AES-T1200 HT from Trust-Hub, HT\#3 and its variant, HT\#4, are denial-of-service (DoS) attacks targeting AES-128 encryption. Both attacks function as encryption bypasses. In HT\#3, the attack is triggered by a specific plaintext input. HT\#4 is more generalized, activating when any plaintext from a predefined set of values is observed. Once either HT is activated, the plaintext is passed to the output unencrypted.

\begin{figure*}[t]
\centering
\begin{tabular}{ccc}
\\
\subfloat\{\includegraphics[width=52mm]{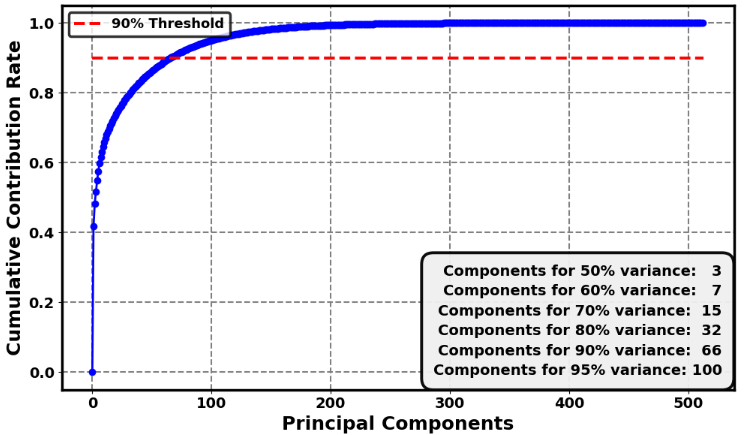}
{(1)}
&
\subfloat{\includegraphics[width=55mm]{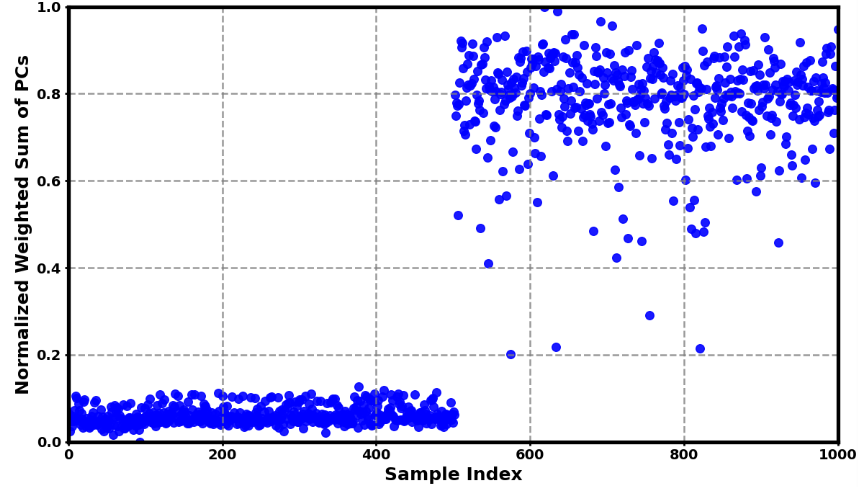}}
{(2)}
&
\subfloat{\includegraphics[width=31mm]{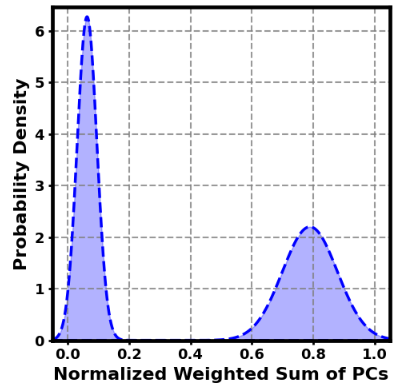}}
{(3)}
\\
\end{tabular}
\caption {(1) Cumulative PCA variance for the information-leakage hardware Trojan (HT\#1); (2) normalized weighted sum of PCs of AES-128 under HT activation; and (3) BGMM-identified Gaussian clusters revealing distinct leakage behavior.} 
\label {fig:HT1}
\vspace{-0.2in}
\end{figure*}

\begin{figure*}[t]
\centering
\begin{tabular}{ccc}
\\
\subfloat{\includegraphics[width=51mm]{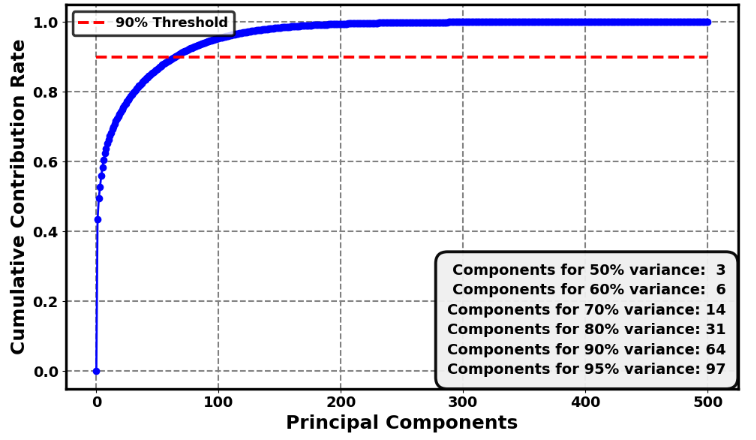}}
{(1)}
&
\subfloat{\includegraphics[width=55mm]{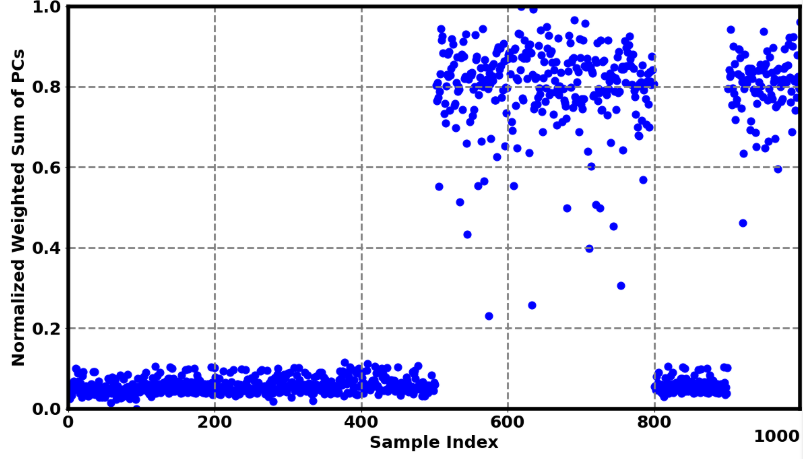}}
{(2)}
&
\subfloat{\includegraphics[width=32mm]{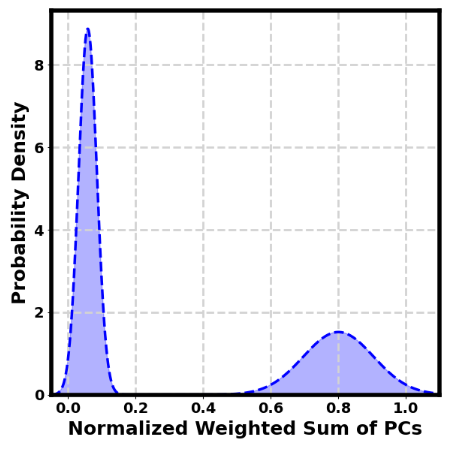}}
{(3)}
\\
\end{tabular}
\vspace{-0.1in}
\caption {(1) PCA variance for the multi-trigger information-leakage hardware Trojan (HT\#2); (2) normalized weighted sum of PCs for dual activation events; and (3) BGMM distributions confirming intermittent HT activation.} 
\vspace{-0.3in}
\label {fig:HT2}
\end{figure*}

To determine the optimal EM probe placement for analyzing HT effects, a systematic methodology was used to generate an EM emissions heatmap of the FPGA surface. The FPGA was mapped to an $8 \times 8$ grid, and the EM probe was placed at each of the 64 coordinates to measure peak-to-peak signal amplitude. By comparing the EM emissions with and without AES encryption, we could identify areas of highest cryptographic activity. As shown in Figure \ref{fig:heatmap}, the AES implementation (case b) resulted in a significant increase in peak-to-peak amplitude (60–80 mV), far exceeding the minor fluctuations observed without AES (32–37 mV) (case a). This contrast revealed the FPGA's most electromagnetically active region, which led to the final probe placement at the center, angled towards a corner, for subsequent experiments. 

Figure \ref{fig:SampleEMs} presents three representative EM traces collected after identifying the active regions of the FPGA through heatmap analysis.
A part of AES-128 core is placed in grid X3Y7, while grid X0Y1 corresponds to an inactive region without any circuit implementation.
As shown, the EM trace from X0Y1 exhibits weak activity with low amplitude, whereas the traces from X3Y7 display stronger signals with significantly higher amplitudes, reflecting active switching behavior in the AES region. 
An EM trace from X3Y7 corresponds to the case where HT\#1 is inactive with an input plaintext of 400, while another trace from the same region corresponds to the case where HT\#1 is activated with the input plaintext is 700.
Following the proposed detection flow shown in Figure \ref{fig:EMHTDetFlow}, each EM signal is transformed into a time–frequency representation using the CWT to generate its corresponding scalogram.
The scalograms illustrated in Figure \ref{fig:SampleEMs} demonstrate how HT activation alters the time–frequency structure of the EM emissions, visibly changing the scalogram’s shape and energy distribution. 

\section{Experimental Evaluations and Results}
\label{ExperimentalEvaluationsandResults}

Figure \ref{fig:HT1}(2) shows the normalized weighted sum of PCs for AES-128 encryption across 1,000 distinct plaintext inputs, reconstructed using 66 principal components corresponding to a variance threshold of 90\%, as determined in Figure \ref{fig:HT1}(1).
In this experiment, HT\#1 activates upon detecting the sequence of plaintexts 500–503 and remains active thereafter.
Figure \ref{fig:HT1}(3) presents the two Gaussian distributions identified by BGMM across all inputs — one centered near 0.06 (variance 0.0005) and the other slightly near 0.8 (variance 0.0090) — clearly revealing a shift in EM behavior associated with HT activation.

To assess robustness against multiple activations, HT\#2—derived from HT\#1—was designed to trigger twice at distinct time intervals.
Figure \ref{fig:HT2}(1) identifies 64 principal components corresponding to a variance threshold of 90\%.
Figure \ref{fig:HT2}(2) displays the normalized weighted sum of PCs for AES-128 encryption under HT\#2 activation, while Figure \ref{fig:HT2}(3) shows two distinct BGMM distributions centered at 0.05 and 0.8 with variances of 0.0004 and 0.0101, respectively.
These results confirm the proposed method’s ability to capture HT behavior even when activation occurs intermittently or in separate operational intervals.

\begin{figure*}[t]
\centering
\begin{tabular}{ccc}
\\
\subfloat{\includegraphics[width=51mm]{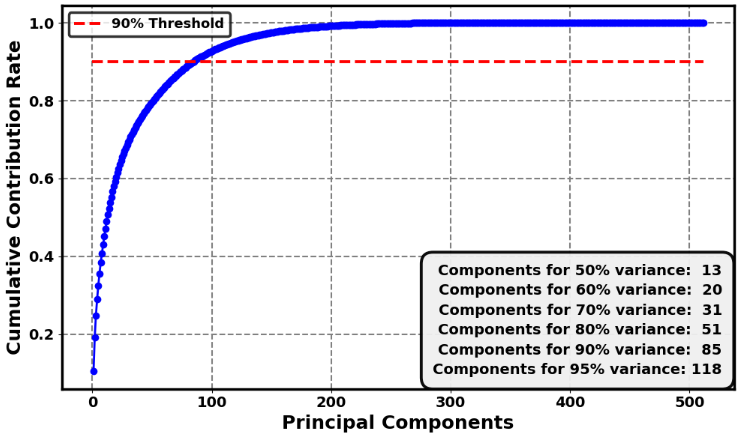}}
{(1)}
&
\subfloat{\includegraphics[width=55mm]{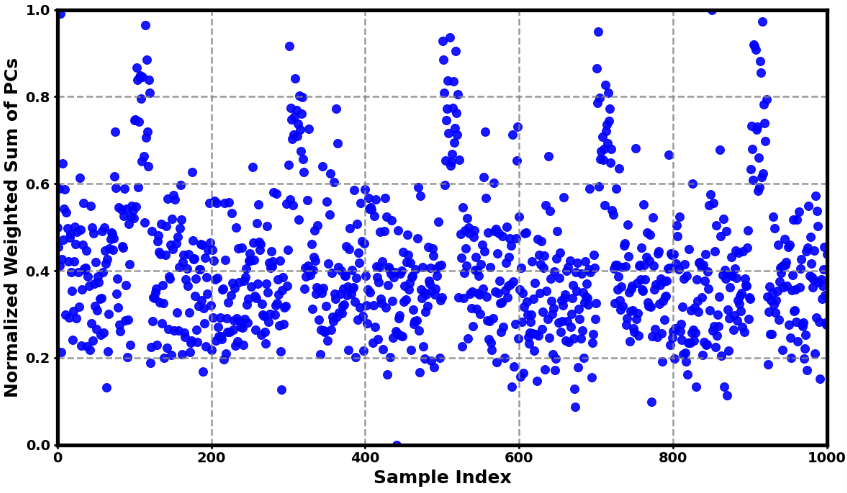}}
{(2)}
&
\subfloat{\includegraphics[width=32mm]{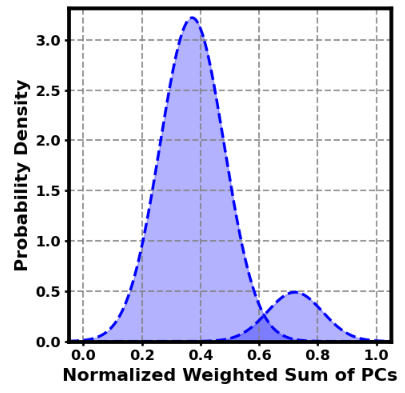}}
{(3)}
\\
\end{tabular}
\vspace{-0.1in}
\caption {(1) PCA variance for the denial-of-service hardware Trojan (HT\#3); (2) normalized weighted sum of PCs under active and inactive states; and (3) BGMM-identified distributions distinguishing subtle activation signatures.} 
\vspace{-0.3in}
\label {fig:HT3}
\end{figure*}

In the case of HT\#3, activation occurs when a specific plaintext value (0x7) is applied during subsequent encryption rounds 7, 100–120, 300–320, 500–520, 700–720, and 900–920 out of 1,000 total rounds.
Once triggered, the HT disables encryption, producing a denial-of-service (DoS) condition in which plaintext is directly forwarded to the output.
When inactive, AES-128 operates normally.
Figure \ref{fig:HT3}(1) identifies 85 principal components at a variance threshold of 90\%, while Figure \ref{fig:HT3}(2) shows the normalized weighted sum of PCs under both active and inactive states across 1,000 rounds.
Figure \ref{fig:HT3}(3) depicts the two BGMM distributions centered near 0.36 (variance 0.0089) and 0.7 (variance 0.0133).
Although these distributions overlap slightly, BGMM successfully distinguishes them, demonstrating the method’s sensitivity to subtle activation signatures.

\begin{figure*}[t]
\centering
\begin{tabular}{ccc}
\\
\subfloat{\includegraphics[width=51mm]{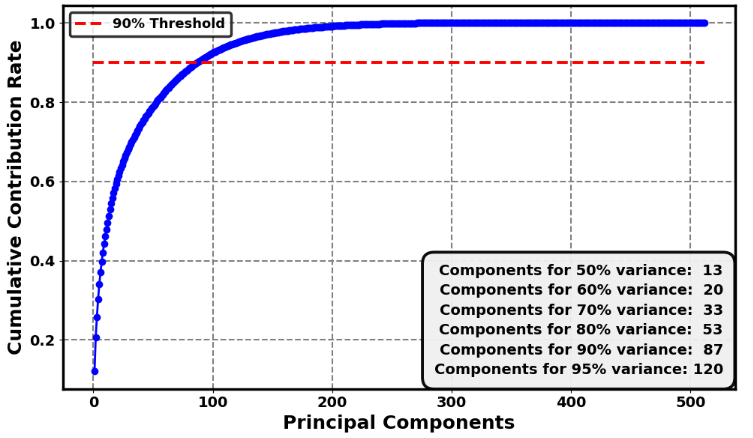}}
{(1)}
&
\subfloat{\includegraphics[width=55mm]{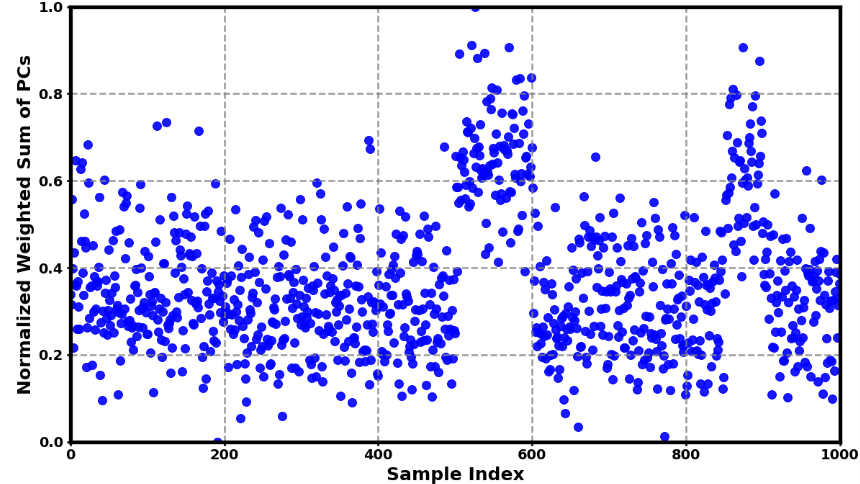}}
{(2)}
&
\subfloat{\includegraphics[width=32mm]{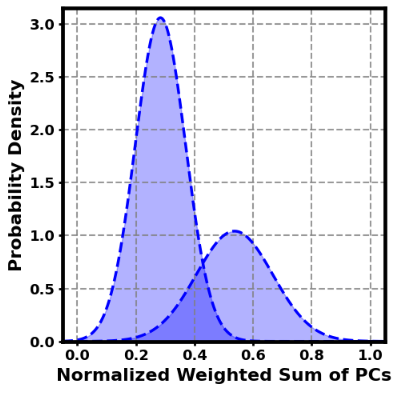}}
{(3)}
\\
\end{tabular}
\vspace{-0.1in}
\caption {(1) PCA variance for the generalized denial-of-service hardware Trojan (HT\#4); (2) EM responses for two representative plaintext inputs; and (3) BGMM-derived Gaussian clusters validating the method’s adaptability across HT types.} 
\vspace{-0.2in}
\label {fig:HT4}
\end{figure*}

Finally, to generalize the approach, HT\#4—inspired by HT\#3—was designed to trigger upon detecting predefined plaintext inputs.
Figure \ref{fig:HT4}(1) shows 87 principal components selected under a variance threshold of 90\%.
Figure \ref{fig:HT4}(2) illustrates the EM responses for two representative plaintext inputs, and Figure \ref{fig:HT4}(3) presents the two Gaussian distributions identified by BGMM, centered at 0.3 and 0.6, with variances of 0.0072 and 0.0130, respectively.
These results validate the proposed flow’s adaptability across various HT behaviors and activation conditions.

Table~\ref{tlb:ConfidenceMetrics} summarizes the quantitative metrics derived 
from the BGMM across all tested configurations under a 90\% PCA variance threshold. 
For the HT-free AES-128 baseline, 
a single Gaussian cluster dominates ($\pi_{max} = 1.0$, $D = 0$), 
with $\Delta BIC < 0$ and $\beta_{post} = 0$, 
confirming the statistical uniformity of EM emissions in the absence of HT activity.

In contrast, the insertion of triggered HTs 
introduces distinct secondary clusters with measurable separation. 
For information-leakage HTs (HT\#1 and HT\#2), 
the BGMM reveals nearly balanced mixture weights ($\pi_{min} \approx 0.4$--$0.5$) 
and large cluster separations ($D > 10$), 
accompanied by high $\Delta BIC$ values ($>2700$). 
Corresponding $\beta_{post}$ levels (0.39--0.49) 
indicate strong confidence in anomaly presence and reflect 
the persistent activation of leakage circuitry once triggered.

For denial-of-service HTs (HT\#3 and HT\#4), 
the mixture weights are less balanced ($\pi_{min} \approx 0.2-0.3$), 
and separations are smaller ($D \approx 3$), 
consistent with shorter activation intervals and smaller EM impact. 
Nonetheless, $\Delta BIC$ remains significantly above 
the decision threshold ($\Delta BIC \geq 10$), 
confirming detectable distributional deviation. 
The $\beta_{post}$ values (0.21--0.35) reflect moderate yet 
reliable confidence in detecting these transient activation events.

Overall, the results demonstrate that 
the proposed reference-free EM-based approach can differentiate 
between HT-free and HT-affected behavior 
with high statistical confidence. 
The combination of $\beta_{\text{post}}\ge0.21$ and $\Delta\mathrm{BIC}\gg10$ forms the primary statistical evidence for HT presence, while $D$ serves as supporting geometric context that characterizes the degree of separability across activation mechanisms.
\textit{Due to space constraints, we report representative results corresponding 
to the 90\% PCA variance threshold. Similar trends were consistently observed 
across other thresholds, confirming 
the robustness and stability of the proposed detection framework.}


The results demonstrate that the proposed EM-based framework
reliably detects diverse HT behaviors without requiring a golden
reference or design knowledge. The consistent separation
metrics ($D$), large $\Delta\mathrm{BIC}$ values, and stable posterior
confidences ($\beta_{\text{post}}$) across HT types and PCA thresholds
confirm that the method is design-agnostic and robust to
implementation details. Moreover, because EM traces are processed
sequentially and maintain statistical consistency across
different observation windows, the approach naturally extends to
\textit{runtime or in-field monitoring}, where continuous analysis of
incoming EM data enables timely detection of emerging anomalies
and supports real-time trust assurance in deployed systems.

\begin{table*}
\centering
\begin{tabular}{|p{0.04\textwidth}|c|p{0.12\textwidth}|c|p{0.15\textwidth}|p{0.15\textwidth}|}
\hline
\multirow{3}{*}{\textbf{Name}} & \multirow{3}{*}{\textbf{Distributions}} & \multicolumn{3}{|c|}{\textbf{Anomaly Determination}} & \multirow{3}{*}{\textbf{Confidence Level}} \\\cline{3-5}
 & \multirow{2}{*}{\textbf{Dominant Cluster}}&\multicolumn{2}{|c|}{\textbf{Secondary Cluster}} & \\\cline{4-5}
& & & \textbf{Mixture Weight} & \textbf{Cluster Separation} & \\\hline

HT-free & 
\parbox[c]{0.18\textwidth}{\centering
\includegraphics[width=1.2\linewidth, height=3cm, keepaspectratio]{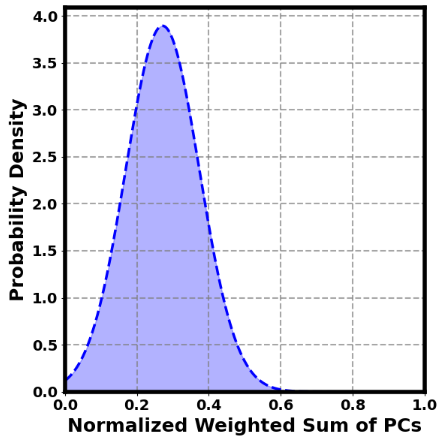}}&
\parbox[c]{0.30\textwidth}{
$\pi_{\text{max}}$ = 1.0000\\
$\pi_{\text{min}}$ = 0.0000\\}&
0 &
\parbox[c]{0.30\textwidth}{
D = 0.0000\\
$\mu_{\text{dom}}$ = [0.2710]\\ $\mu_{\text{anom}}$ = [0]\\
$\Sigma_{\text{dom}}$ = [0.0105]\\ $\Sigma_{\text{anom}}$ = [0]\\
$\Sigma_{\text{avg}}$ = [0]}&
\parbox[c]{0.22\textwidth}{
$\alpha_{\text{post}}$= 1\\
$\beta_{\text{post}}$ = 0\\
$\Delta \text{BIC}$ = -18.72}\\\hline

HT\#1&
\parbox[c]{0.18\textwidth}{\centering
\includegraphics[width=1.2\linewidth, height=3cm, keepaspectratio]{new_image_paper_LEAKAGE_INFORMATION_UPDATED,_DISTROBUTION.png}}&
\parbox[c]{0.30\textwidth}{
$\pi_{\text{max}}$ = 0.5045\\
$\pi_{\text{min}}$ = 0.4955\\}&
49.55\%&
\parbox[c]{0.30\textwidth}{
D = 10.5148\\
$\mu_{\text{dom}}$ = [0.0622]\\ $\mu_{\text{anom}}$ = [0.7889]\\
$\Sigma_{\text{dom}}$ = [0.0005]\\ $\Sigma_{\text{anom}}$ = [0.0090]\\
$\Sigma_{\text{avg}}$ = [0.0048]}&
\parbox[c]{0.22\textwidth}{
$\alpha_{\text{post}}$ = 0.5050\\
$\beta_{\text{post}}$ = 0.4950\\
$\Delta \text{BIC}$ = 2760.0996}\\\hline

HT\#2&
\parbox[c]{0.18\textwidth}{\centering
\includegraphics[width=1.2\linewidth, height=3cm, keepaspectratio]{new_image_paper_LEAKAGE_INFORMATION_manipulate_UPDATED,_DISTROBUTION.png.png}}&
\parbox[c]{0.30\textwidth}{
$\pi_{\text{max}}$ = 0.6024\\
$\pi_{\text{min}}$ = 0.3976\\}&
39.76\%&
\parbox[c]{0.30\textwidth}{
D = 10.2905\\
$\mu_{\text{dom}}$ = [0.0577]\\ $\mu_{\text{anom}}$ = [0.8018]\\
$\Sigma_{\text{dom}}$ = [0.0004]\\ $\Sigma_{\text{anom}}$ = [0.0101]\\
$\Sigma_{\text{avg}}$ = [0.0052]}&
\parbox[c]{0.22\textwidth}{
$\alpha_{\text{post}}$ = 0.6030\\
$\beta_{\text{post}}$ = 0.3970\\
$\Delta \text{BIC}$ = 3250.90}\\\hline

HT\#3&
\parbox[c]{0.18\textwidth}{\centering
\includegraphics[width=1.2\linewidth, height=3cm, keepaspectratio]{new_image_paper_DOS_UPDATED,_DISTROBUTION.png}}&
\parbox[c]{0.30\textwidth}{
$\pi_{\text{max}}$ = 0.7887\\
$\pi_{\text{min}}$ = 0.2113\\}&
21.13\%&
\parbox[c]{0.30\textwidth}{
D = 3.2240\\
$\mu_{\text{dom}}$ = [0.3536]\\ $\mu_{\text{anom}}$ = [0.6931]\\
$\Sigma_{\text{dom}}$ = [0.0089]\\ $\Sigma_{\text{anom}}$ = [0.0133]\\
$\Sigma_{\text{avg}}$ = [0.0111]}&
\parbox[c]{0.22\textwidth}{
$\alpha_{\text{post}}$ = 0.7885\\
$\beta_{\text{post}}$ = 0.2115\\
$\Delta \text{BIC}$ = 193.9755}\\\hline
HT\#4&
\parbox[c]{0.18\textwidth}{\centering
\includegraphics[width=1.2\linewidth, height=3cm, keepaspectratio]{new_image_paper_DOS_manipulate_UPDATED,_DISTROBUTION.png}}&
\parbox[c]{0.30\textwidth}{
$\pi_{\text{max}}$ = 0.6473\\
$\pi_{\text{min}}$ = 0.3527\\}&
35.27\%&
\parbox[c]{0.30\textwidth}{
D = 2.9801\\
$\mu_{\text{dom}}$ = [0.2839]\\ $\mu_{\text{anom}}$ = [0.5832]\\
$\Sigma_{\text{dom}}$ = [0.0072]\\ $\Sigma_{\text{anom}}$ = [0.0130]\\
$\Sigma_{\text{avg}}$ = [0.0101]}&
\parbox[c]{0.22\textwidth}{
$\alpha_{\text{post}}$ = 0.6477\\
$\beta_{\text{post}}$ = 0.3523\\
$\Delta \text{BIC}$ = 135.3569}\\\hline
\end{tabular}
\caption{BGMM-based statistical confidence parameters for HT detection at 90\% PCA variance threshold.}
\vspace{-0.3in}
\label{tlb:ConfidenceMetrics}
\end{table*}

\section{Related Work}
\label{RelatedWork}

Hardware Trojan detection has progressed through structural,
side-channel, and learning-based approaches. Among these,
EM side-channel analysis is particularly appealing for
post-silicon assurance due to its non-invasive nature and
fine spatial resolution. However, most EM-based techniques
still rely on golden reference data or labeled training sets,
restricting their use in modern distributed manufacturing
flows where no trusted baseline is available.

Golden-reference techniques, such as those by Jap \emph{et al.}~\cite{jap2016supervised} 
and He \emph{et al.}~\cite{he2017chipfree}, compare side-channel signatures 
from suspect chips against known-good exemplars. 
While effective in controlled test conditions, 
these methods require prior knowledge of the original circuit or chip layout, 
which is impractical in modern globalized semiconductor supply chains. 
Other side-channel works \cite{jain2021survey} 
use correlated power or path-delay features for HT identification, 
but they rely on access to internal design data or precise timing models 
that may not be available after fabrication.

To overcome data dependency, recent studies have leveraged deep and transfer learning. 
Sun \emph{et al.}~\cite{sun2022transfer} proposed 
an EM-image classification approach using pre-trained convolutional neural networks, 
while Lee \emph{et al.}~\cite{lee2024robust} introduced a robust autoencoder 
for anomaly detection in hardware security traces. 
Although these methods demonstrate improved adaptability, 
they still require either labeled datasets or architecture-specific retraining, 
making them unsuitable for black-box post-silicon analysis. 
Moreover, they generally lack interpretable metrics to quantify detection confidence.

Zhang \emph{et al.} \cite{Zhang2019Hardware} proposed a HT method 
using electromagnetic leakage with PCA and Mahalanobis distance. 
While effective, it depends on a golden reference chip and suffers 
from PCA’s linearity and noise sensitivity, limiting robustness in complex environments. 
Similarly, Liao and Meng \cite{Liao2024Hardware} introduced a model 
using electromagnetic side-channel leakage with PCA for feature reduction, 
clustering, and enhanced KNN for detection. 
Like \cite{Zhang2019Hardware}, their approach also relies on a golden model, 
which may not always be available, and faces additional limitations 
such as potential information loss, sensitivity 
to noise, parameter tuning, and limited scalability.

In contrast, the proposed work introduces 
a fully reference-free and design-agnostic EM-based HT detection framework. 
By combining CWT, PCA, and BGMM, 
the method performs unsupervised probabilistic inference directly on raw EM data. 
Unlike prior art, it quantifies statistical separability and confidence 
through metrics such as $\Delta BIC$, $\alpha_{post}$, and $\beta_{post}$, 
enabling interpretable detection without any labeled or golden data. 
Furthermore, the approach’s adaptability to sliding-window and time-sequential analysis 
allows extension to runtime or in-field HT monitoring—representing 
a fundamental shift from laboratory-only validation 
to scalable, continuous trust assurance in microelectronic systems.

\section{Conclusion}
\label{Conclusion}
This work introduced a reference-free EM side-channel framework for detecting triggered HTs in fabricated integrated circuits without any golden model, labeled data, or design access. The proposed flow combines CWT, CNN feature extraction, PCA, and BGMM to perform unsupervised probabilistic anomaly detection with interpretable statistical confidence metrics. Experimental validation on AES-128 circuits embedded with various HTs demonstrated clear statistical separation between HT-free and HT-activated cases, supported by posterior confidence indicators ($\alpha_{post}$, $\beta_{post}$), cluster separations ($D$), and $\Delta BIC$ values. Overall, this methodology establishes a scalable and interpretable foundation for post-silicon and operational trust assurance, where EM side-channel signatures themselves become self-verifying indicators of hardware integrity.
While this work focuses on AES-128 as a representative case, the proposed framework is adaptable to a broad range of digital designs. Future efforts will extend evaluation to non-cryptographic and mixed-signal systems to further validate scalability and trigger diversity.

\centering
\textbf{\\Acknowledgement\\}
This work is supported by the Office of Naval Research under Grant N000142312131.

\bibliographystyle{IEEEtran} 
\bibliography{references}

@article{chesebrough2017trusted,
  author    = {Dave Chesebrough},
  title     = {Trusted Microelectronics: A Critical Defense Need},
  journal   = {National Defense Magazine},
  year      = {2017},
  month     = {October},
  url       = {https://www.nationaldefensemagazine.org/articles/2017/10/31/trusted-microelectronics-a-critical-defense-need},
  note      = {Accessed: February 2025}
}

@techreport{bis2023microelectronics,
  author    = {{U.S. Department of Commerce, Bureau of Industry and Security}},
  title     = {Assessment of the Status of the U.S. Microelectronics Industrial Base},
  institution = {U.S. Department of Commerce},
  year      = {2023},
  month     = {December},
  url       = {https://www.bis.gov/media/documents/section-9904-report-final-20231221.pdf},
  note      = {Section 9904 Report to the President}
}

@article{TahghighArXiv2026,
  author  = {Mahsa Tahghigh and Hassan Salmani},
  title   = {Reference-Free Spectral Analysis of {EM} Side-Channels for Always-on Hardware {T}rojan Detection},
  journal = {arXiv preprint arXiv:2601.20163},
  year    = {2026},
  url     = {https://arxiv.org/abs/2601.20163}
}

@article{cta3980,
author = {Tahghigh, Mahsa and Shiri, Nabiollah},
title = {A new ripple carry adder structure based on a swing-boosted full adder for concurrent error correction in low-resolution pipeline analog-to-digital converters},
journal = {International Journal of Circuit Theory and Applications},
volume = {52},
number = {9},
pages = {4741-4754},
doi = {https://doi.org/10.1002/cta.3980},
year = {2024}
}

@INPROCEEDINGS{11014346,
  author={Elahi, Mehdi and Elshamy, Mohamed R. and Badawy, Abdel-Hameed and Fazeli, Mahdi and Patooghy, Ahmad},
  booktitle={2025 26th International Symposium on Quality Electronic Design (ISQED)}, 
  title={Matter: Multi-Stage Adaptive Thermal Trojan for Efficiency \& Resilience Degradation}, 
  year={2025},
  volume={},
  number={},
  pages={1-8},
  doi={10.1109/ISQED65160.2025.11014346}}

@ARTICLE{8952724,
  author={Huang, Zhao and Wang, Quan and Chen, Yin and Jiang, Xiaohong},
  journal={IEEE Access}, 
  title={A Survey on Machine Learning Against Hardware Trojan Attacks: Recent Advances and Challenges}, 
  year={2020},
  volume={8},
  number={},
  pages={10796-10826},
  doi={10.1109/ACCESS.2020.2965016}
}

@ARTICLE{9310331,
  author={Hu, Wei and Chang, Chip-Hong and Sengupta, Anirban and Bhunia, Swarup and Kastner, Ryan and Li, Hai},
  journal={IEEE Transactions on Computer-Aided Design of Integrated Circuits and Systems}, 
  title={An Overview of Hardware Security and Trust: Threats, Countermeasures, and Design Tools}, 
  year={2021},
  volume={40},
  number={6},
  pages={1010-1038},
  doi={10.1109/TCAD.2020.3047976}
}

@article{Xiao2016Hardware,
author = {Xiao, Kan and Forte, Domenic and Jin, Y. and Karri, R. and Bhunia, S. and Tehranipoor, Mark},
year = {2016},
month = {05},
pages = {1-23},
title = {Hardware Trojans: Lessons Learned after One Decade of Research},
volume = {22},
journal = {ACM Transactions on Design Automation of Electronic Systems},
doi = {10.1145/2906147}
}

@article{Etherington2019,
  author    = {Thomas R Etherington},
  title     = {Mahalanobis distances and ecological niche modelling: correcting a chi-squared probability error},
  journal   = {PeerJ},
  year      = {2019},
  doi       = {10.7717/peerj.6678}
}

@article{pan2022survey,
  author  = {Z. Pan and P. Mishra},
  title   = {A Survey on Hardware Vulnerability Analysis Using Machine Learning},
  journal = {IEEE Access},
  volume  = {10},
  pages   = {49508--49527},
  year    = {2022},
  doi     = {10.1109/ACCESS.2022.3173287}
}

@inproceedings{jain2021survey,
  author  = {A. Jain and Z. Zhou and U. Guin},
  title   = {Survey of Recent Developments for Hardware Trojan Detection},
  booktitle = {2021 IEEE International Symposium on Circuits and Systems (ISCAS)},
  address = {Daegu, Korea},
  pages   = {1--5},
  year    = {2021},
  doi     = {10.1109/ACCESS.2022.3173287}
}

@inproceedings{hanindhito2019hardware,
  author  = {B. Hanindhito and Y. Kurniawan},
  title   = {Hardware Trojan Design and Its Detection using Side-Channel Analysis on Cryptographic Hardware AES Implemented on FPGA},
  booktitle = {2019 International Conference on Electrical Engineering and Informatics (ICEEI)},
  address = {Bandung, Indonesia},
  pages   = {191--196},
  year    = {2019}
}

@inproceedings{vashistha2018trojan,
  author  = {N. Vashistha and H. Lu and Q. Shi and M. T. Rahman and H. Shen and D. L. Woodard},
  title   = {Trojan Scanner: Detecting Hardware Trojans with Rapid SEM Imaging Combined with Image Processing and Machine Learning},
  booktitle = {ISTFA Conference Proceedings},
  year    = {2018},
  doi     = {10.31399/asm.cp.istfa2018p0256}
}

@inproceedings{tahghigh2024gmm,
  author  = {M. Tahghigh and H. Salmani},
  title   = {Detecting Hardware Trojans in Manufactured Chips Without Reference: A {GMM}-Based Approach},
  booktitle = {IEEE/ACM International Conference on Computer-Aided Design (ICCAD '24)},
  address = {New York, NY, USA},
  year    = {2024},
  doi     = {10.1145/3676536.3689919}
}

@inproceedings{ghimire2023quantum,
  author  = {A. Ghimire and A. A. Hossain and N. P. Bhatta and F. Amsaad},
  title   = {Identification and Localization of Quantum Electromagnetic Fields of Hardware Trojan Attacks Using QDM-Based Unsupervised Deep Learning},
  booktitle = {2023 IEEE Physical Assurance and Inspection of Electronics (PAINE)},
  address = {Huntsville, AL, USA},
  pages   = {1--7},
  year    = {2023},
  doi     = {10.1109/PAINE58317.2023.10317976}
}

@article{he2017chipfree,
  author  = {J. He and Y. Zhao and X. Guo and Y. Jin},
  title   = {Hardware Trojan Detection Through Chip-Free Electromagnetic Side-Channel Statistical Analysis},
  journal = {IEEE Transactions on Very Large Scale Integration (VLSI) Systems},
  volume  = {25},
  number  = {10},
  pages   = {2939--2948},
  year    = {2017}
}

@inproceedings{jap2016supervised,
  author  = {D. Jap and W. He and S. Bhasin},
  title   = {Supervised and unsupervised machine learning for side-channel based trojan detection},
  booktitle = {Proc. IEEE 27th Int. Conf. Appl.-Specific Syst., Archit. Processors (ASAP)},
  pages   = {17--24},
  year    = {2016}
}

@article{gubbi2023tutorial,
  author  = {K. I. Gubbi and B. S. Latibari and A. Srikanth and T. Sheaves and S. A. Beheshti-Shirazi and S. M. PD and S. Rafatirad and A. Sasan and H. Homayoun and S. Salehi},
  title   = {Hardware Trojan Detection Using Machine Learning: A Tutorial},
  journal = {ACM Transactions on Embedded Computing Systems},
  volume  = {22},
  number  = {3},
  year    = {2023},
  doi     = {10.1145/3579823}
}

@inproceedings{gubbi2022survey,
  author  = {K. I. Gubbi and S. A. Beheshti-Shirazi and T. Sheaves and S. Salehi and S. M. PD and S. Rafatirad and A. Sasan and H. Homayoun},
  title   = {Survey of Machine Learning for Electronic Design Automation},
  booktitle = {Proc. Great Lakes Symposium on VLSI (GLSVLSI 2022)},
  pages   = {513--518},
  year    = {2022}
}

@article{lee2024robust,
  author  = {D. Lee and J. Lee and Y. Jung and J. Kauh and T. Song},
  title   = {Robust Hardware Trojan Detection Method by Unsupervised Learning of Electromagnetic Signals},
  journal = {IEEE Transactions on Very Large Scale Integration (VLSI) Systems},
  volume  = {32},
  number  = {12},
  pages   = {2327--2340},
  year    = {2024}
}

@article{sun2022transfer,
  author  = {S. Sun and H. Zhang and X. Cui and L. Dong and X. Fang},
  title   = {Electromagnetic Side-Channel Hardware Trojan Detection Based on Transfer Learning},
  journal = {IEEE Transactions on Circuits and Systems II: Express Briefs},
  volume  = {69},
  number  = {3},
  pages   = {1742--1746},
  year    = {2022},
  doi     = {10.1109/TCSII.2021.3110954}
}

@article{Zhang2019Hardware,
  author={L. Zhang and Y. Dong and J. Wang and C. Xiao and D. Ding},
  title={A Hardware Trojan Detection Method Based on the Electromagnetic Leakage},
  journal={China Communications},
  volume={16},
  number={12},
  pages={100--110},
  year={2019},
  doi={10.23919/JCC.2019.12.007} 
}

@inproceedings{Liao2024Hardware,
  author={L. Liao and F. Meng},
  title={Hardware Trojan Detection and Identification Using Electromagnetic Side-Channel Leakage},
  booktitle={2024 9th International Conference on Intelligent Computing and Signal Processing (ICSP)},
  pages={328--333},
  year={2024},
  doi={10.1109/ICSP62122.2024.10743629}
}

@String{Computing = "Computing" }

\end{document}